\documentclass[sigconf,screen]{acmart}
\usepackage[utf8]{inputenc}
\usepackage{hyperref}
\usepackage{soul}
\hypersetup{
  colorlinks   = true, 
  urlcolor     = blue, 
  linkcolor    = blue, 
  citecolor    = red 
}

\copyrightyear{2022} 
\acmYear{2022} 
\setcopyright{acmlicensed}
\acmConference[SPLC '22]{26th International Systems and Software Product Line Conference - Volume A}{September 12-16, 2022}{Graz, Austria}
\acmBooktitle{26th International Systems and Software Product Line Conference - Volume A (SPLC '22), September 12-16, 2022, Graz, Austria}

\title{A Benchmark for Active Learning of Variability-Intensive Systems}

\author{Shaghayegh Tavassoli}
\email{sh.tavassoli@ut.ac.ir}
\orcid{0000-0002-2474-7390}
\affiliation{%
  \institution{University of Tehran}
  \city{Tehran}
  \country{IR}
}

\author{Carlos Diego N. Damasceno}
\email{d.damasceno@cs.ru.nl}
\orcid{0000-0001-8492-7484}
\affiliation{%
  \institution{Radboud University Nijmegen}
    \city{Nijmegen}
  \country{NL}
}

\author{Mohammad Reza Mousavi}
\orcid{0000-0002-4869-6794}
\email{mohammad.mousavi@kcl.ac.uk}
\affiliation{%
  \institution{King's College London}
  \city{London}
  \country{UK}
  \postcode{WC2R 2LS}
}

\author{Ramtin Khosravi}
\orcid{0000-0001-6393-0959}
\email{r.khosravi@ut.ac.ir}
\affiliation{%
  \institution{University of Tehran}
  \city{Tehran}
  \country{IR}
}

\begin{document}

\begin{abstract}
Behavioral models are the key enablers for behavioral analysis of Software Product Lines (SPL), including testing and model checking. Active model learning comes to the rescue when family behavioral models are non-existent or outdated. A key challenge on active model learning is to detect commonalities and variability efficiently and combine them into concise family models. Benchmarks and their associated metrics will play a key role in shaping the research agenda in this promising field and provide an effective means for comparing and identifying relative strengths and weaknesses in the forthcoming techniques. In this challenge, we seek benchmarks to evaluate the efficiency (e.g., learning time and memory footprint) and effectiveness (e.g., conciseness and accuracy of family models) of active model learning methods in the software product line context. These benchmark sets must contain the structural and behavioral variability models of at least one SPL. Each SPL in a benchmark must contain products that requires more than one round of model learning with respect to the basic active learning $L^{*}$ algorithm. Alternatively, tools supporting the synthesis of artificial benchmark models are also welcome. 
\end{abstract}


\maketitle

\section{Introduction}

Variability in a software system enables mass production and customization, particularly in the context of diverse and evolving requirements \cite{DBLP:conf/wicsa/GurpBS01,DBLP:journals/tse/GalsterWTMA14}. 
To maximize these opportunities, one needs to analyze variability-intensive systems in a structured manner by considering variabilities and commonalities as important characteristics of such software systems.
Family-based models are an enabler for such a structured analysis approach. 
Different types of family models have been used to analyze SPL functionality, including structural models \cite{kang1990feature,DBLP:conf/re/SchobbensHT06} and behavioral models \cite{DBLP:journals/tse/ClassenCSHLR13,DBLP:journals/ese/DamascenoMS21}. A type of structural model widely used in software product line engineering (SPLE) research is feature models \cite{kang1990feature,DBLP:conf/re/SchobbensHT06}. Structural models can be used to annotate traditional behavioral models to represent the behavior of a product line \cite{DBLP:journals/tse/ClassenCSHLR13,DBLP:journals/ese/DamascenoMS21}. Featured Transition System (FTS) \cite{DBLP:journals/tse/ClassenCSHLR13} and Featured Finite State Machines (FFSM) \cite{DBLP:conf/facs2/FragalSM16,DBLP:journals/cj/FragalSMT19} are two examples of behavioral variability models for expressing the state-based behavior of families of software products. The FTS model extends traditional labelled transition systems with Boolean expressions for annotating the transitions. Likewise, the FFSM model is an extension of Mealy Finite State Machines (FSM) \cite{gill1962introduction} that uses Boolean expressions on states and transitions as presence conditions  \cite{DBLP:journals/cj/FragalSMT19,DBLP:conf/splc/DamascenoMS19}. 

Active model learning is a process in which a model of software behavior is learned by presenting different inputs and observing the output \cite{DBLP:journals/iandc/Angluin87,DBLP:journals/cacm/Vaandrager17,DBLP:conf/ifm/DamascenoMS19}. Model learning can be used in the behavioral analysis of an SPL, for example in model-based testing and model checking \cite{DBLP:journals/ese/DamascenoMS21}.  When applied to variability-intensive systems, it is expected that the interactions with different variants or products will help to identify commonalities and variability. 
Adaptive model learning is a model learning approach in which previous models are reused to learn a new model, so that the speed of model learning can be increased \cite{DBLP:conf/ifm/DamascenoMS19}. In an SPL, the number of valid products may be exponential relative to the number of features \cite{DBLP:journals/ese/DamascenoMS21,DBLP:journals/tse/BergerSLWC13}. However, due to the commonalities of SPL products, it is possible to reuse existing models during the model learning of an SPL. Therefore, it may be possible to reduce the total cost of learning the SPL model by using these common features \cite{DBLP:journals/ese/DamascenoMS21}.

Active model learning is performed using two types of queries, namely membership queries (MQ) and equivalence queries (EQ). The number of EQs and MQs and the length of these queries can affect the efficiency of model learning \cite{DBLP:journals/iandc/Angluin87,DBLP:journals/cacm/Vaandrager17,DBLP:conf/ifm/DamascenoMS19}. 
To evaluate the model learning methods in the context of SPLs, it is necessary to have a benchmark set that can clearly show how efficiently and accurately different learning methods perform. In a suitable benchmark, it should be feasible to evaluate each learning cost metric (described in Section \ref{sec:metrics}). Therefore, such benchmarks should contain products that require multiple rounds to be learned using usual non-adaptive methods to accurately show the effect of learning methods on models with different characteristics.

\section{Problem Definition}

In this challenge, participants are asked to provide a benchmark set that can be used for evaluating model learning methods in variability-intensive systems. The benchmark set must contain at least one SPL. Each SPL in the benchmark set must contain at least one product which requires more than one iteration for model learning using the $L^{*}$ algorithm. The variability models and the behavioral models of these SPLs must be provided in the benchmark set. The characteristics that result in a multi-round model learning process should also be explained in the solution report. Alternatively, a tool for generating artificial SPL models can be provided as a solution. In this case, the tool must be able to generate random FFSMs or families of FSMs by taking the variability model and the input/output alphabets as input arguments.

\section{Background}

In this section, some of the terms and definitions used in this paper are briefly explained. First, some well-known notations for structural and behavioral modeling of SPLs are described. Then, the model learning process and the metrics used to evaluate model learning methods are explained.

\subsection{Feature Model}

A feature model \cite{kang1990feature,DBLP:conf/re/SchobbensHT06} is a variability model that demonstrates the structure of an SPL as a hierarchical representation of SPL features. A feature model shows the commonality and variability of SPL products using a tree representation, implying the relationship between the features such that whenever a (non-root) feature exists in a configuration, its parent feature must exist in that configuration too. In this representation, mandatory features are specified using a solid circle. Optional features are represented using an empty circle. Alternative features are demonstrated as sibling nodes in the tree which are specified using an arc drawn between the edges of all options. If a product contains the parent feature, only one of the alternative features must be present in that product \cite{kang1990feature,DBLP:conf/re/SchobbensHT06,DBLP:journals/ese/DamascenoMS21}. Feature dependencies can also be expressed using simple cross-tree constraints (i.e., \textit{requires} and \textit{excludes} relations) or using boolean logic expressions defined over the feature set \cite{DBLP:conf/splc/Batory05}.

\begin{figure}[!ht]
\centering\includegraphics[scale=0.6]{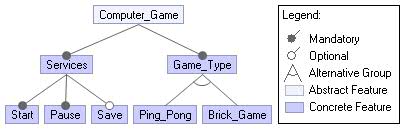}
\caption{The feature model of a simple computer game product line, inspired by \cite{DBLP:conf/facs2/FragalSM16}}
\label{fig:figure_1}
\end{figure}

The feature model of a simple computer game is shown in Figure \ref{fig:figure_1} (taken from \cite{DBLP:conf/facs2/FragalSM16} with minor modifications). In this figure, \textit{Save} is an optional feature. \textit{Services}, \textit{Start}, \textit{Pause} and \textit{Game$\_$Type} are mandatory features. In each valid product, exactly one feature from $\{ \mathit{Ping}\_\mathit{Pong},\mathit{Brick}\_\mathit{Game}\}$ is present. Using a feature constraint, one can concisely specify one or more product configurations by expressing a boolean expression over the features. For example, the feature constraint $\neg\mathit{Save}$ specifies the set of all product configurations which does not include the \textit{Save} feature \cite{kang1990feature,DBLP:conf/re/SchobbensHT06}.

\subsection{Featured Behavioral Models}

Various methods have been proposed for behavioral modeling of SPLs. Many of these notations are extensions of Finite State Machines (FSM) \cite{gill1962introduction} or Labeled Transition Systems (LTS) \cite{DBLP:books/daglib/0020348} that can also incorporate features. Some notable examples are Featured Finite State Machines (FFSM) \cite{DBLP:conf/facs2/FragalSM16,DBLP:journals/cj/FragalSMT19} and Featured Transition Systems (FTS) \cite{DBLP:conf/icse/ClassenHSLR10,DBLP:journals/tse/ClassenCSHLR13}. Given that most model learning methods aim at learning FSMs \cite{DBLP:journals/cacm/Vaandrager17}, in this section, we explain FFSM notation, which is an FSM-based featured behavioral model \cite{DBLP:conf/facs2/FragalSM16,DBLP:journals/cj/FragalSMT19}. The same concepts can be used to come up with benchmarks based on other family-based models such as FTSs. 

\begin{figure}[!ht]
\centering\includegraphics[scale=0.4]{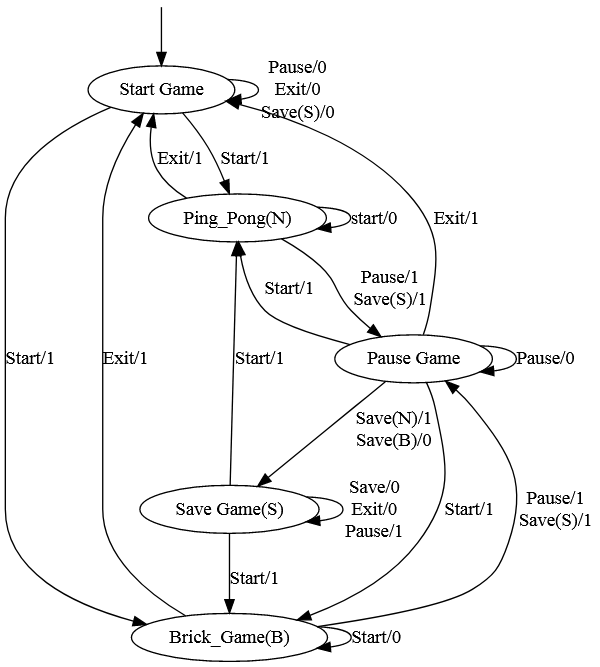}
\caption{The FFSM of a simple computer game product line, inspired by  \cite{DBLP:conf/facs2/FragalSM16}}
\label{fig:figure_2}
\end{figure}

A Finite State Machine (FSM) is a behavioral model and is defined as $M=\langle S,s_{0},I,O,\delta ,\lambda\rangle$. In this definition, $S$ is the set of FSM states and $s_{0}$ is its initial state. The set of input and output alphabets are represented by $I$ and $O$, respectively. The transition function is denoted by $\delta$ which determines what the next state is when the FSM is in state $s\in S$ and the input $a\in I$ is presented. The output function is denoted by $\lambda$ which maps each pair of a state and an input to an output. An FSM is deterministic if for each state and input, there exists at most one transition \cite{DBLP:journals/cacm/Vaandrager17,DBLP:conf/ifm/DamascenoMS19}.

A Featured Finite State Machine (FFSM) can be defined as an FSM whose states and transitions are labeled with feature constraints. An FFSM is defined as $(F,\Lambda , C,c_{0},Y,O,\Gamma )$. In this definition, $F$ is the set of features and $\Lambda$ is the set of valid configurations. The set of conditional states is represented by $C$. A conditional state is defined as $(s,b)$ where $s$ is a state and $b$ is a feature constraint. The initial conditional state is denoted by $c_{0}$ and is defined as $(s_{0},true)$. The set of conditional inputs is denoted by $Y$ and the set of outputs is represented by $O$. Finally, $\Gamma$ specifies the set of conditional transitions. A conditional transition is defined as a tuple $(s_{i},y,o,s_{j})$ where $s_{i}$ and $s_{j}$ are conditional states, $y$ is a conditional input and $o$ is an output. An FFSM can be projected into the FSM of any given valid product using the product derivation operator \cite{DBLP:conf/facs2/FragalSM16,DBLP:journals/cj/FragalSMT19,DBLP:journals/ese/DamascenoMS21}. 

\begin{figure}[!ht]
\centering\includegraphics[width=0.8\linewidth]{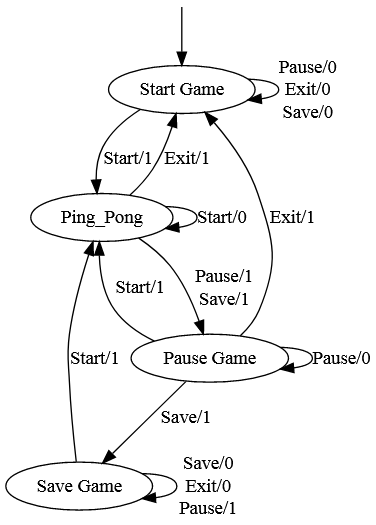}
\caption{The FSM of a product derived from the computer game product line, inspired by \cite{DBLP:conf/facs2/FragalSM16}}
\label{fig:figure_3}
\end{figure}

The FFSM model of the computer game example described in 3.1 is shown in Figure \ref{fig:figure_2}. In this figure, the feature constraints of the states and inputs are written in capital letters in parentheses. The \textit{Save} feature is shown with the letter $S$, the $\mathit{Ping}\_\mathit{Pong}$ feature with the letter $N$ and the $\mathit{Brick}\_\mathit{Game}$ feature with the letter $B$. Figure \ref{fig:figure_3} shows the FSM of a product containing $\mathit{Ping}\_ \mathit{Pong}$ and \textit{Save} features derived from the FFSM in Figure \ref{fig:figure_2} (Figures \ref{fig:figure_2} and \ref{fig:figure_3} are taken from \cite{DBLP:conf/facs2/FragalSM16} with minor modifications).

In this challenge, participants are asked to include, if possible, the FSMs of all valid SPL products in the benchmark set. If the number of valid products is large, a featured behavioral model should be provided (instead of the FSMs of all products). In this case, it is preferred to provide a tool that can be used to derive the FSM of any given valid configuration using the featured behavioral model.

\subsection{Model Learning}

Model learning \cite{DBLP:journals/cacm/Vaandrager17} is a method for constructing a state machine that represents the behavior of a software system. During the active model learning process, different input sequences are presented to the system under learning (SUL) and its output is observed. Based on the observed results, a hypothesis about the SUL behavior (the FSM language) is constructed. In the $L^{*}$ algorithm \cite{DBLP:journals/iandc/Angluin87} proposed by Dana Angluin, two types of queries are used to learn the behavior of the SUL (a deterministic FSM). Using a membership query (MQ), it is asked what the SUL output is for a given input sequence. This type of queries is used in the hypothesis construction phase of learning. Using an equivalence query (EQ) it is asked whether hypothesis $H$ about the SUL language is correct. If the hypothesis is correct, the learning process will be terminated. If the hypothesis is incorrect, a counterexample is provided in which the result of $H$ differs from the SUL language. Equivalence queries are used in the hypothesis validation phase of model learning 
\cite{DBLP:journals/iandc/Angluin87,DBLP:journals/cacm/Vaandrager17,DBLP:conf/ifm/DamascenoMS19}.

During the model learning process, the query results are stored in an observation table. An observation table $OT$ can be defined as a triple $(S,E,T)$, where $S$ is a set of prefixes which is prefix-closed. A prefix-closed set of strings is a set that contains all the prefixes of all of its members. In the definition of the observation table, $E$ is a set of suffixes. Assuming $s\in S$ and $e\in E$, $T[s,e]$ is the last output of the SUL when the input sequence is $s.e$ (i.e., $s$ concatenated with $e$). In the $L^{*}$ algorithm, when a counterexample $c$ is returned, $c$ and its prefixes are added to $S$ in the observation table and the new values of $T$ are calculated using MQs. This refined observation table will be used in the next iteration of the algorithm \cite{DBLP:conf/sfm/SteffenHM11}. While the $L^{*}$ algorithm was originally designed for inferring deterministic finite automata, we focus on the variant tailored to Mealy machines \cite{DBLP:conf/fm/ShahbazG09,DBLP:conf/ifm/DamascenoMS19}.

\subsection{Metrics}
\label{sec:metrics}

The time complexity of model learning algorithms depends on the number of EQs and MQs used in the learning process. In each iteration of $L^{*}$, one equivalence query (conjecture) is used. Therefore, the number of iterations (rounds) can be considered as an important learning cost metric \cite{DBLP:journals/iandc/Angluin87,DBLP:journals/cacm/Vaandrager17,DBLP:conf/sfm/SteffenHM11}. Another parameter that affects the time complexity of learning algorithms, is the length of EQs and MQs. The total number of symbols used in MQs and in the implementation of EQs can be used as another cost metric \cite{DBLP:journals/cacm/Vaandrager17}. Also, the number of reset operations can affect the cost of learning \cite{DBLP:conf/ifm/DamascenoMS19}. To evaluate the efficiency of model learning methods, it is necessary to be able to compare the values of these metrics in different learning methods. A suitable benchmark set and its associated metrics can be very useful when evaluating the model learning methods.

\section{The Challenge}

In this challenge, participants are asked to provide a benchmark set that can be used to evaluate model learning methods in the software product line context.

Alternatively, a tool support for generating artificial SPLs, i.e., \textit{random FFSMs} or \textit{family of FSMs randomly generated} for a given variability model; is also welcome.

\subsection{Content of the Challenge Solution}

The benchmark set must contain the following items for at least one SPL:
\begin{enumerate}
\item \textbf{Variability model:} As a variability model, the SPL feature model must be provided. The feature model must be in the XML format of the FeatureIDE \cite{DBLP:journals/scp/ThumKBMSL14} library.

\item \textbf{Behavioral model:} To demonstrate the behavior of the SPL, one of the following methods must be used:
\begin{itemize}
\item If possible, the FSM of each valid configuration of the SPL must be provided in the format of a dot file.
\item If the number of valid products of an SPL is large, the following items must be provided:
\begin{enumerate}
\item A featured behavioral model representing the behavior of the SPL
\item A derivation tool that can derive the FSM of any given valid configuration using the featured behavioral model of the SPL (the derived FSMs must be in dot format)
\end{enumerate}
\end{itemize}
\end{enumerate}

\noindent
{\bf Random FFSM/FSM model generator.} Alternatively, participants can submit methods to generate artificial SPLs in the format of random families of FSMs (or FFSMs), with their respective implementations. A sample of artificial models produced with their method must be provided according to the requirements previously discussed. 

The methods submitted must allow users to define the following parameters:
\begin{enumerate}
    \item The variability source (i.e., feature model)
    \item The randomness source (i.e., seed)
    \item The number of states of the generated FSMs/FFSM
    \item The set of input/output symbols (i.e., I/O alphabets)
\end{enumerate}

\subsection{Characteristics of the Benchmark Set}

This benchmark set should be applicable to evaluate the efficiency, conciseness, and accuracy of model learning methods in the SPL context. Using this benchmark set, it should be possible to compare model learning cost metrics, such as EQs, MQs, the number of rounds, conciseness, and accuracy.
The metrics should show some reduction in the cost of learning the family model (e.g., in the number of MQs, EQs, or the total number of symbols) with respect to learning the individual products. 
Two other important metrics to be evaluated using the benchmark-set are the conciseness \cite{DBLP:conf/splc/DamascenoMS19} of the family model and its accuracy \cite{DBLP:journals/ese/DamascenoMS21}. The learned family model is supposed to merge the commonalities into common states and transitions and hence, ought to be much smaller than the sum of the size of the individual products. Moreover, the learned family model should be accurate in the sense that all product behaviors derived from the learned family model should be equivalent to the products of the original product-line under learning. To enable measuring these two metrics, we need to have access to the original (possibly hand-crafted) models of the product lines and a tool for deriving product models from the family model.

Each SPL in this benchmark set must contain at least one configuration (SPL product) that requires more than one round for its model to be learned using the $L^{*}$ algorithm. To measure the learning cost metrics, the learning experiments have to be performed using LearnLib \cite{DBLP:conf/fase/RaffeltS06} library version 0.16.0\footnote{\url{ https://github.com/LearnLib/learnlib/tree/learnlib-0.16.0}}. 

The non-adaptive model learning can be performed using ExtensibleLStarMealyBuilder class from the LearnLib library \footnote{\url{http://learnlib.github.io/learnlib/maven-site/0.14.0/apidocs/de/learnlib/algorithms/lstar/mealy/ExtensibleLStarMealyBuilder.html}} using the following parameters:
\begin{itemize}
\item The set of initial prefixes is $S=\{ \epsilon\}$.
\item The set of initial suffixes is $E=I$, where $I$ is the input alphabet.
\item The values of other learning parameters are optional. These parameters include:
\begin{itemize}
\item The equivalence oracle type
\item The counterexample handling method
\item The observation table closing strategy
\item Whether caching is used or not
\end{itemize}
The assumed values for these parameters should be explained in the solution report.
\end{itemize}

Participants are also asked to explain the reasons why some product models require more than one round to be learned. The characteristics of such products can be explained in the solution report.

The detailed information for providing the benchmark set is available on GitHub\footnote{\url{https://github.com/damascenodiego/splc22challenge_dataset_learning}}. In this repository, there is a Java class called \textit{AutomataLearning} with a code example for experimenting with the LearnLib framework \footnote{\url{https://learnlib.de/}} for active model learning \cite{DBLP:conf/fase/RaffeltS06}. This class takes an FSM file in dot format as an input. The model of this FSM is learned using the $L^{*}$ algorithm and the values of the learning cost metrics are printed.

\section{Summary}

In this challenge, participants are asked to provide a benchmark set that can be used for evaluating model learning methods in variability-intensive systems. This benchmark set should be applicable for comparing the efficiency of different model learning methods in the SPL context. 

The provided benchmark set must contain at least one SPL, in which at least one of its products requires more than one round for its model to be learned using the $L^{*}$ method. The benchmark set should contain the variability models and the behavioral models of the SPLs. Alternatively, a tool for generating families of artificial FSMs/FFSMs can be provided, in combination with a sample of random behavioral variability models. The participants are also asked to explain the reasons why some product models require more than one round for model learning. The submitted solutions will be evaluated using the requirements and the experimental setup explained in section 4.2. The set of accepted solutions can be used in the future as benchmarks for evaluating model learning methods in variability-intensive systems.

\begin{acks}
Mohammad Reza Mousavi has been partially supported by
the UKRI Trustworthy Autonomous Systems Node in Verifiability, Grant Award
Reference EP/V026801/1.
\end{acks}

\bibliographystyle{ACM-Reference-Format}
\bibliography{main}
\end{document}